\documentclass[lettersize,journal]{IEEEtran}

\IEEEoverridecommandlockouts
\def\BibTeX{{\rm B\kern-.05em{\sc i\kern-.025em b}\kern-.08em
   T\kern-.1667em\lower.7ex\hbox{E}\kern-.125emX}}

\usepackage{amsmath,amssymb,amsfonts}
\usepackage{amsmath,amsfonts}
\usepackage{algorithmic}
\usepackage{algorithm}
\usepackage{array}
\usepackage[caption=false,font=normalsize,labelfont=sf,textfont=sf]{subfig}
\usepackage{textcomp}
\usepackage{stfloats}
\usepackage{url}
\usepackage{verbatim}
\usepackage{graphicx}
\usepackage{cite}
\usepackage{xcolor}
\usepackage{ifthen}
\usepackage{calligra}
\usepackage{listings}
\usepackage{rotating}
\usepackage{tcolorbox}

\usepackage{tikz}
\usetikzlibrary{tikzmark}
\usepackage{enumitem}
\usepackage{authblk}

\usepackage{atbegshi}

\newcommand{\roundbox}[1]{\tikzmarknode[fill=cyan,fill opacity=0.3,draw=green!60!black,thick,rounded corners,inner sep=2pt,text opacity=1]{round}{#1}}
\newboolean{showcomments}
\setboolean{showcomments}{true}
\ifthenelse{\boolean{showcomments}}
 { \newcommand{\mynote}[2]{
      \fbox{\bfseries\sffamily\scriptsize#1}
        {\small$\blacktriangleright$\textsf{\emph{#2}}$\blacktriangleleft$}}}
        { \newcommand{\mynote}[2]{}}
\definecolor{DarkOrange}{rgb}{0.8,0.3,0.0}
\definecolor{DarkCyan}{rgb}{0.0, 0.55, 0.55}
\definecolor{DarkCyel}{rgb}{1.0, 0.49, 0.0}
\definecolor{yellow-green}{rgb}{0.6, 0.8, 0.2}

\newcolumntype{?}{!{\vrule width 1pt}}

\def\BibTeX{{\rm B\kern-.05em{\sc i\kern-.025em b}\kern-.08em
    T\kern-.1667em\lower.7ex\hbox{E}\kern-.125emX}}
\usepackage{balance}
\title{A Systematic Survey on Debugging Techniques for Machine Learning Systems}

\author[1]{Thanh-Dat Nguyen}
\author[1]{Haoye Tian}
\author[1]{Bach Le}
\author[1]{Patanamon Thongtanunam}
\author[2]{Shane McIntosh}

\affil[1]{School of Computing and Information Systems, The University of Melbourne \protect\\ \texttt{thanhdat.nguyen@student.unimelb.edu.au}, \texttt{haoye.tian@unimelb.edu.au}, \texttt{bach.le@unimelb.edu.au},
\texttt{patanamon.t@unimelb.edu.au}
}
\affil[2]{David R. Cheriton School of Computer Science, University of Waterloo\protect\\ 
\texttt{shane.mcintosh@uwaterloo.ca}}

\begin{document}

\maketitle

\begin{abstract}
Machine learning (ML) and deep learning (DL) in general have been widely applied as state-of-the-art techniques in a wide range of domains.
Like traditional software, ML systems are also prone to faults during development, which can degrade their performance. However, debugging ML software (i.e., the detection, localization and fixing of faults) poses unique challenges compared to traditional software largely due to the probabilistic nature and heterogeneity of its development process.
Over the past decade, various methods have been proposed for testing, diagnosing, and repairing ML systems.
However, the big picture informing important research directions that really address the dire needs of developers is yet to unfold, leaving several key questions unaddressed: (1) What faults have been targeted in the ML debugging research that fulfill developers' needs in practice? (2) How are these faults addressed? (3) What are the challenges in addressing the yet untargeted faults? Given the rapid development of ML and their increasingly significant impact, a timely review aligning research and practice in ML debugging is needed to inform future research in this area.

In this paper, we conduct a systematic study of debugging techniques for machine learning systems. We first collect technical papers focusing on debugging components in machine learning software. We then map these papers to a taxonomy of faults to assess the current state of fault resolution identified in existing literature. Subsequently, we analyze which techniques are used to address specific faults based on the collected papers. This results in a comprehensive taxonomy that aligns faults with their corresponding debugging methods.
Finally, we examine previously released transcripts of interviewing developers to identify the challenges in resolving unfixed faults.
Our analysis reveals that only 48\% of the identified ML debugging challenges have been explicitly addressed by researchers, while 46.9\% remain unresolved or unmentioned. In real-world applications, we found that 52.6\% of issues reported on GitHub and 70.3\% of problems discussed in interviews are still unaddressed by research in ML debugging. The study identifies 13 primary challenges in ML debugging, categorized into data-related, framework-related, and conceptual/resource-related issues. Among these, the most common challenges include domain-specific data processing, hard-to-use frameworks, and difficulties in understanding models and training processes. Our findings highlight a significant gap between the identification of ML debugging issues and their resolution, particularly in practical settings. This underscores the need for more targeted research and development efforts to address these persistent challenges in ML system debugging.

\end{abstract}

\section{Introduction}

Machine learning (ML), especially deep learning, is now widely recognized as the state-of-the-art in many domains. Due the rapid development and adoptions of ML, traditional software development processes have gradually shifted to a data-driven approach~\cite{Amershi2019}. Along this transition, ML development has also become different from traditional software engineering: ML software developments are generally heterogeneous~\cite{Panichella2021} (i.e., consisting of multiple components such as data, training algorithm, model architecture, etc.), while ML models typically remain black-box. The heterogeneous and black-box nature of ML systems make it challenging to identify, localize, and fix bugs in the systems (referred to as ML debugging). 


Faults in ML systems can manifest from any ML pipeline components~\cite{Humbatova2020taxonomyofrealfault, Croft2023} and identifying faults and affected components that need to be investigated has proved to be difficult~\cite{deepdiagnosis, ma2018deepmutation, hu2019deepmutationplus, deepfd, Panichella2021}. Recently, a survey by Zhang et al. summarized numerous research efforts on testing ML systems to reveal faults, for example, testing correctness, robustness, etc~\cite{Zhang2022}. Other work attempted to study bugs in ML systems~\cite{islam2019comprehensivednnbugs} and subsequently created a practical fine-grained taxonomy of real faults in ML systems~\cite{Humbatova2020taxonomyofrealfault}. This taxonomy showed the landscape of existing faults and the components in which the faults can often manifest. To create the taxonomy, a comprehensive interview and survey with practitioners and developers have been conducted to summarize faults and problems that have been frequently encountered with ML systems in practice. Additionally, GitHub issues were also considered to augment more data besides the survey. In a similar vein, Islam et al.~\cite{islam2019comprehensivednnbugs} studied the types of bugs in ML systems and their root causes by iterating through StackOverflow and GitHub, deriving more coarse-grained bug categories such as structural vs non-structural bugs. 

Despite the promising progress of these works on faults detection and categorization, faults and testing of ML systems are only a part of the larger landscape of ML debugging that further requires analyzing buggy ML systems, localizing buggy components, and fixing them. Debugging ML systems presents unique challenges due to the opacity of complex models, intricate data pipelines, and the non-deterministic nature of many ML algorithms that complicate the debugging process~\cite{Zhang2022}. Traditional software debugging methods, such as those that use code instrumentation and breakpoints, are often inadequate to address these complexities due to the distinctive characteristics of ML workflows~\cite{Amershi2019}. As ML software is becoming mainstream with an increasingly significant impact, a deeper understanding of the larger landscape of existing ML debugging research is required to promptly inform future research in this area.


In this work, we provide a systematic survey of the current ML debugging landscape, identifying the relevance between research and practical needs in this area. By doing so, our study offers practical implications on which areas more research efforts should be spent on to address real pressing challenges in practice. 

We seek to answer the following questions:
                        
\begin{enumerate}[wide, labelwidth=!, labelindent=0pt]
\item \textbf{What faults have been targeted by ML debugging in the literature? Is resolving those faults in harmony with developers' need in practice?} We collect papers in the literature on debugging techniques and link the types of faults targeted in these papers to the taxonomy of real faults proposed by Humbatova et al.~\cite{Humbatova2020taxonomyofrealfault}. We consider this taxonomy because it is fine-grained and its data are publicly released, which was created in a comprehensive way by a survey and interview with professional developers and by real ML bugs in GitHub repositories. By this categorization, we identify both (1) which faults have or have not been targeted by existing research in ML debugging techniques and (2) whether resolving those types of faults fulfills developers' need in practice.
\item \textbf{How are the faults being addressed?} We systematically construct our own taxonomy of debugging techniques and analyze the connection between the taxonomy of faults and our taxonomy of debugging techniques. The alignment between the taxonomies allows us to comprehensively categorize ML debugging methods, linking them to real ML faults that need to be resolved in practice.
 \item \textbf{What are the challenges in targeting the remaining untargeted faults?} For the faults that have not been targeted by current ML debugging methods, we identify the challenges of fixing these faults by studying the interview transcription provided by the taxonomy of faults~\cite{Humbatova2020taxonomyofrealfault}. We conducted a thematic analysis, following common practices~\cite{Braun01012006, Cruzes2011ThematicSynthesis}, to discover the common challenges that were mentioned during the interview for machine learning faults debugging.
\end{enumerate}

In summary, our study reveals that only 48\% of the identified ML debugging challenges have been explicitly addressed by researchers, while 46.9\%
remain unresolved or unmentioned. In real-world applications, we found that 52.6\% of issues reported on GitHub and 70.3\% of problems discussed in interviews with practitioners are still unaddressed by ML debugging research. We identify 13 primary challenges in ML debugging, including data-related, framework-related, and conceptual/resource-related issues. Our
findings highlight a significant gap between the identification of ML debugging issues and their resolution in practical settings. This underscores the need for more targeted research and development efforts to address these persistent challenges in debugging ML systems.

In the following sections, to clearly explain the scope of the paper and related concepts, we give details of background in machine learning system faults and testing in Section~\ref{sec:background}. Section~\ref{sec:methodology} explains our methodology to systematically conduct this survey, describing the alignments between the taxonomy of faults and the taxonomy of debugging techniques in ML systems. Section~\ref{sec5:gaps} identifies the gaps between ML faults and their current resolutions. Section~\ref{sec6:implications} provides implications for future research in ML debugging, followed by Section~\ref{sec7:conclusion} that concludes.
\section{Background and Motivation}\label{sec:background}
In this section, we first give details on the machine learning development process in Section~\ref{subsec:ml_dev_background}. Each step in the development process constructs a corresponding component of machine learning pipelines, which we describe in Section~\ref{subsec:ml_component_background}. Section~\ref{subsec:ml_component_testing_background} describes how these components are tested with the corresponding testing techniques and testing criteria. Finally, we describe the motivation of our paper in Section~\ref{sec:testing_to_debugging}.
 
\subsection{Machine Learning Development Process}\label{subsec:ml_dev_background}
Machine learning development generally consists of nine stages~\cite{Amershi2019}: model requirements, data collection, data cleaning, data labeling, feature engineering, model training, model evaluation, model deployment, and model monitoring.
In case a fault is introduced in any of these stages, the effect can be propagated, leading to a sequence of faults~\cite{Androutsopoulos2014ErrorPropagation} on the subsequent stages of the development pipeline. 

\emph{Model Requirements:} In this step, the developer chooses model architecture as well as a specific framework and implements the model. Choosing the wrong version of the framework, suboptimal architecture, or having the wrong implementation will lead to a sub-optimal performance in terms of correctness or crashes~\cite{islam2019comprehensivednnbugs, Zhang2018}.

\emph{Data Collection:} Bugs during this phase, such as collecting wrong data or missing out on important data sources, can introduce bias or incompleteness, including biased or unfair model~\cite{Chakraborty2021}.

\emph{Data Cleaning:} If data is not cleaned properly, the model might be trained on noise or outliers, which could adversely impact its performance. Mistakes in handling missing values, removing duplicates, or normalizing data can introduce significant errors.

\emph{Data Labeling:} Incorrect labels can mislead the model during training. Even a small fraction of mislabeled data can result in poor model generalization, especially in cases where the dataset is already limited.

\emph{Feature Engineering:} Creating inappropriate features or missing out on important features can lead to models that are either overfitting (too complex) or underfitting (too simple). Bugs here might cause the model to focus on irrelevant features or wrongly processed features, leading to decrease in accuracy or crashes of the model.

\emph{Model Training:} Bugs in this phase, such as choosing an inappropriate algorithm, wrong hyperparameters, or training for an insufficient number of epochs, can hinder the model's ability to learn from the data effectively.

\emph{Model Evaluation:} If the evaluation metrics or test data are flawed, one might end up with a false sense of the model's performance. Bugs here could lead to deploying models that are not ready or overlooking models that are performing well.

\emph{Model Deployment:} Deploying a model with bugs can lead to wrong predictions in a live environment. This might not only reduce user trust but could also have significant consequences, especially in sensitive areas like healthcare or finance.

\emph{Model Monitoring:} Proper model monitoring is essential to ensure models maintain optimal performance. Given inadequate monitoring, the system may fail to detect issues such as model drift or degradation. As a result, outdated models could continue making suboptimal predictions, leading to negative outcomes.

\subsection{Machine Learning Pipeline Components}\label{subsec:ml_component_background}
During the development process, the following artifacts are produced.

\emph{Collected Data}: The raw data gathered from various sources, which serves as the input for the machine learning pipeline. This data may be structured, semi-structured, or unstructured.

\emph{Preprocessed Data}: The output of the data preprocessing stage, where the collected data is cleaned, transformed, and normalized to ensure consistency and compatibility with the machine learning model.

\emph{Labeled Data}: The preprocessed data that has been annotated with target labels or values, which is used for supervised learning tasks. Labeling can be done manually, semi-automatically, or through crowdsourcing.

\emph{Feature Sets}: The engineered features extracted from the preprocessed data, which serve as the input to the machine learning model. Feature engineering involves selecting, creating, and transforming relevant features to improve model performance.

\emph{Model Architecture}: The structure and design of the machine learning model, including the choice of algorithm, number and type of layers, activation functions, and other hyperparameters.

\emph{Trained Model}: The model that has been trained on the labeled data using the selected architecture and hyperparameters. The trained model encapsulates the learned patterns and relationships from the data.

\emph{Evaluation Metrics}: The quantitative measures used to assess the performance of the trained model, such as accuracy, precision, recall, F1-score, or mean squared error, depending on the type of task and dataset.

\emph{Testing Data}: A separate dataset that is used to evaluate the generalization ability of the trained model. Testing data should be representative of the real-world scenarios in which the model will be deployed.

\emph{Deployed Model}: The trained model that is integrated into a production environment, such as a web service, mobile app, or embedded system, to make predictions or decisions based on new, unseen data.

These components form the building blocks of a machine learning system and are the targets for testing and debugging techniques discussed in the following sections.

\subsection{Machine Learning Testing}\label{subsec:ml_component_testing_background}
In machine learning systems, various components or combinations of components are tested to ensure the correctness, robustness, and reliability of the overall system. In detail, according to a recent survey by Zhang et al.~\cite{Zhang2022}, the testing targets can be categorized as follows.

\emph{Correctness}: Correctness testing aims to ensure that the machine learning system produces the expected outputs for given inputs. This involves testing the model's accuracy, precision, recall, and other relevant metrics on various datasets, including corner cases and edge cases~\cite{Barr2015, Breck2017}. Techniques such as cross-validation, model assertions, and metamorphic testing are used to verify the correctness of the model's predictions~\cite{Xie2011, Segura2018}.

\emph{Efficiency}: Efficiency testing focuses on evaluating the performance and resource utilization of the machine learning system. This includes measuring the training and inference time, memory usage, and computational complexity of the model~\cite{Amershi2019}. Techniques such as profiling, benchmarking, and complexity analysis are used to identify bottlenecks and optimize the system's efficiency~\cite{Justus2018}.

\emph{Fairness}: Fairness testing aims to detect and mitigate biases and discriminatory behavior in machine learning systems. This involves testing the model's performance across different subgroups and ensuring that it does not exhibit disparate treatment or disparate impact~\cite{Galhotra2017, Bellamy2018}. Techniques such as statistical parity, equality of opportunity, and counterfactual fairness are used to assess and improve the model's fairness~\cite{Dwork2012, Kusner2017}.

\emph{Reliability}: Reliability testing evaluates the machine learning system's ability to perform consistently and handle various operating conditions. This includes testing the model's robustness to noise, missing data, and distribution shifts, as well as its fault tolerance and error handling capabilities~\cite{Guo2017, Hendrycks2019}. Techniques such as stress fuzzing are used to assess the system's reliability~\cite{Tian2018}.

\emph{Security}: Security testing aims to identify and mitigate vulnerabilities and adversarial attacks on machine learning systems. This involves testing the model's resistance to adversarial examples, data poisoning, and model inversion attacks~\cite{Goodfellow2015, Papernot2018}. Techniques such as adversarial training, input validation, and secure aggregation are used to enhance the system's security~\cite{Madry2018, Bonawitz2017}.

\subsection{From Testing to Debugging}\label{sec:testing_to_debugging}
While testing is essential for identifying bugs, faults, or defects in machine learning (ML) systems, it is only the first step in the debugging process. Debugging consists of further steps such as locating, understanding, and resolving these issues. Debugging is critical for ensuring the reliability and trustworthiness of ML systems, as it enables developers to uncover the root causes of problems and implement effective solutions.

However, debugging ML systems presents unique challenges. The opacity of complex models, intricate data pipelines, and the non-deterministic nature of many ML algorithms complicate the debugging process~\cite{Zhang2022}. Traditional software debugging techniques, such as code instrumentation and breakpoint debugging, are often inadequate for addressing these complexities due to the distinctive characteristics of ML workflows~\cite{Amershi2019}.

Given these challenges, there is a pressing need for specialized debugging techniques tailored to ML systems. A comprehensive survey of these debugging methods is crucial to understand the current landscape, identify gaps, and guide future research. By systematically aligning debugging and testing techniques with the types of faults identified in Humbatova et al.’s taxonomy~\cite{Humbatova2020taxonomyofrealfault}, our work provides a structured overview of how faults are detected and fixed in ML environments. This survey not only highlights existing solutions but also emphasizes the importance of developing advanced debugging strategies to enhance the robustness and reliability of machine learning applications.
\section{Methodology}\label{sec:methodology}
To systematically survey the current state of ML debugging techniques and their practicality, we start with collecting ML debugging papers in the literature following our defined selection criteria. We then create a taxonomy of ML debugging techniques by following common practices in thematic analysis~\cite{Braun01012006, Cruzes2011ThematicSynthesis}. We then align our taxonomy of ML debugging with the taxonomy of real faults created by Humbatova et al.~\cite{Humbatova2020taxonomyofrealfault}. This methodology enables us to explore what have been done in research in ML debugging that target real and pressing challenges in practice. In the following, we describe details of our methodology.

\subsection{Paper Selection Criteria}

To answer our research questions, we collect research papers in the literature on ML debugging (i.e., identification, localization, and fixing of ML faults). We define the following inclusion and exclusion criteria for selecting relevant papers.

\noindent\roundbox{\textbf{Inclusion Criteria}:}
\begin{itemize}
\item \textbf{Peer-reviewed papers.} Only articles that have undergone a rigorous peer-review process and have been published in well-established journals or top-tier conferences are included. To ensure the reputation of the publication venues, we consider those indexed in recognized databases such as Scopus or Web of Science and those that are widely acknowledged within the fields of software engineering, software analytics, and artificial intelligence. This approach ensures that the selected studies meet high standards of quality and reliability.
\item \textbf{Full papers.} Only complete manuscripts are considered, excluding abstracts, posters, short papers, or extended abstracts. This criterion guarantees that all necessary details regarding the methodologies, experiments, and findings are available for comprehensive analysis and synthesis.
\item \textbf{Debugging techniques.} Papers that propose or apply debugging techniques for machine learning systems.
\item \textbf{Performance goals.} Papers that focus on improving the performance of machine learning systems in terms of correctness, safety, security, fairness, and efficiency.
\item \textbf{Debugging ML components.} Papers that achieve performance goals by debugging architectural specifications, data preprocessing, output post-processing, training algorithms, or model weights.
\item \textbf{Real-world applications.} Papers that have been evaluated on real-world faults, for example, the faults have been encountered on StackOverflow, GitHub, Huggingface, Kaggle, or production systems.
\end{itemize}
\roundbox{\textbf{Exclusion Criteria}:}
\begin{itemize}
\item \textbf{Exploratory research papers.} Papers that propose ideas or conduct surveys, but do not propose or apply debugging techniques.
\item \textbf{Machine learning techniques.} Papers that do not propose or apply debugging techniques while directly proposing new training algorithms or model architectures.
\end{itemize}

\subsection{Paper Collection Process}
To collect relevant papers, we follow a systematic search process. We begin by querying the Scopus database, a comprehensive and widely-used indexing service that provides access to a vast collection of peer-reviewed literature across various disciplines. Scopus has been employed in numerous systematic reviews and meta-analyses due to its extensive coverage and reliable indexing standards. For venues, we select top-tier conferences and journals from software engineering, software analytics, and artificial intelligence, we use any combination of (``Machine Learning'', ``Deep Learning'') and (``bug'', ``fault'', ``debug'' and ``repair''. The search query is as follows:

\lstset{
  basicstyle=\ttfamily\small,
  breaklines=true,
  keywordstyle=\bfseries,
  breakatwhitespace=false,
  columns=fullflexible,
  showstringspaces=false,
  frame=single,
  keepspaces=true,
  escapeinside={(*@}{@*)},
  morekeywords={CONF, AND, OR, TITLE-ABS-KEY}
}

\begin{lstlisting}
(*@\textbf{CONF}@*) 
  (ICSE (*@\textbf{OR}@*) ASE (*@\textbf{OR}@*) FSE (*@\textbf{OR}@*) ICML (*@\textbf{OR}@*) 
   ICLR (*@\textbf{OR}@*) IJCAI (*@\textbf{OR}@*) ACL (*@\textbf{OR}@*) NeurIPS (*@\textbf{OR}@*) 
   AAAI (*@\textbf{OR}@*) EMNLP (*@\textbf{OR}@*) ISSTA (*@\textbf{OR}@*) ICST (*@\textbf{OR}@*) 
   TSE (*@\textbf{OR}@*) TOSEM) 
(*@\textbf{AND}@*) 
(*@\textbf{TITLE-ABS-KEY}@*) 
  (("Machine Learning" (*@\textbf{OR}@*) "Deep Learning") 
   (*@\textbf{AND}@*) ("bug" (*@\textbf{OR}@*) "fault" (*@\textbf{OR}@*) "debug" (*@\textbf{OR}@*) "repair"))
\end{lstlisting}

To ensure the comprehensiveness of our search strategy, we performed quality control by verifying that our search string successfully retrieves a set of known relevant papers. We compiled an initial set of seminal papers in the domain of ML debugging from previous surveys and expert recommendations. We then confirmed that these papers were included in the search results generated by our query, thereby validating the effectiveness of our search strategy in capturing relevant literature.

Next, we apply predefined inclusion and exclusion criteria to refine our selection of relevant papers. Before proceeding with the full filtering, we conduct a pilot study on a sample of 60 papers. During this pilot phase, the first and second authors independently apply the criteria and perform cross-validation to ensure the reliability and consistency of the selection process. Our agreement rate in this phase exceeds 93\%. After this pilot phase, we perform selecting papers according to inclusion and exclusion criteria. In cases of disagreement, we engage in discussions until a consensus is reached, ensuring that the final set of included papers is robust and agreed upon. Table~\ref{table:keywords} describes details of the numbers of papers we collected based on the described collection process. In total, from the intial collection of 486 papers, including 365 papers for the open-coding set and 121 papers for the validation set, we carefully filter them via the title and the body of the papers to retain only papers that are indeed addressing ML debugging. This process results in the final collection of 96 papers, including 62 papers in the open-coding set and 34 papers in the validation set. 


\subsection{The Taxonomy of ML Debugging Techniques}\label{subsec:taxonomydebugging}
We follow common practices in thematic analysis~\cite{Braun01012006, Cruzes2011ThematicSynthesis} to create a taxonomy of ML debugging based on the papers collected in previous steps. That is, we divide the set of collected papers into two subsets, including open-coding set and validation set. The open-coding set consists of papers published prior to 2023, while the validation set comprises the remaining papers. We first perform open coding~\cite{Cruzes2011ThematicSynthesis} on the open-coding set to establish the initial taxonomy categories. The open-coding method allows us to flexibly start without a predefined code (topic) list, and then inductively discover and merge codes along the way. Once the preliminary taxonomy has been developed from the open-coding set, we apply it to the validation set to assess its robustness and applicability to newer research. By categorizing the papers in the validation set according to the established taxonomy, we are able to evaluate whether the categories adequately captured the diversity of recent advancements in the field. This process helps us identify any emerging topics or areas that were not previously considered, allowing us to refine and expand the taxonomy as necessary. 

We describe in more details about our thematic analysis in what follows. We first extract the initial codes by identifying keywords and phrases related to debugging techniques in each of the collected papers. This process involves extracting relevant key points that describe the techniques proposed in the papers and identifying the exact keywords within these key points to serve as initial codes. We then iteratively refine these initial codes, allowing themes to emerge through repeated analysis and comparison. This iterative process ensures that the taxonomy accurately reflects the range and nuances of debugging techniques present in the literature. The final taxonomy is presented in the lower part of Figure~\ref{fig:taxonomy_of_alignment}.

Having constructed this taxonomy of ML debugging enables us to answer critical questions such as ``what faults are being fixed, are those practically in dire needs by developers, and how they are fixed?''. We describe the details of each category in the constructed  taxonomy of ML debugging techniques in Section~\ref{subsec:fix}.

\begin{table}
\centering
\caption{Keywords used for searching papers on Scopus database}
\begin{tabular}{ccccc}
\hline
\textbf{S. No.} & \textbf{Keyword} & Hits & Title & Body \\
\hline
1 & [Machine Learning]+bug & 101 & 33 & 14 \\
2 & [Machine Learning]+test & 65 & 20 & 10 \\
3 & [Machine Learning]+fault & 53 & 13 & 2 \\
4 & [Machine Learning]+debug & 20 & 12 & 6 \\
5 & [Machine Learning]+repair & 34 & 12 & 3 \\
6 & [Deep Learning]+bug & 90 & 48 & 18 \\
7 & [Deep Learning]+test & 45 & 26 & 14 \\
8 & [Deep Learning]+fault & 37 & 22 & 8 \\
9 & [Deep Learning]+debug & 19 & 13 & 5 \\
10 & [Deep Learning]+repair & 30 & 13 & 4 \\
\hline
Open Coding & All & 365 & 212 & 62 \\
Validation & All & 121 & 73 & 34 \\ 
\hline
\end{tabular}
\label{table:keywords}
\end{table}

\subsection{Alignment Between the Taxonomy of Fault and the Taxonomy of Debugging Techniques}
As stated previously, we want to explore what is the current state of ML debugging landscape with respect to real ML faults that are in real needs to be resolved in practice. Answering this question requires us to know which ML debugging techniques target which types of real ML faults categorized in~\cite{Humbatova2020taxonomyofrealfault}. To do so, we perform coding (labeling) of each ML debugging paper based on the problem that the paper is targeting and align them with the fault taxonomy in~\cite{Humbatova2020taxonomyofrealfault}. We describe this process below.

\vspace{2mm}\noindent\textbf{Fault-Oriented Categorization of Debugging Techniques.}  
To effectively map each ML debugging paper to corresponding faults, we begin by extracting the key points from each paper. These key points capture the specific problems the papers address. We then cross-reference these problems with the fault codes outlined in the taxonomy of faults by Humbatova et al.~\cite{Humbatova2020taxonomyofrealfault}, assigning each paper to the appropriate code(s). It is important to note that a single paper may be associated with multiple fault codes. 

We note that some papers, however, do not address faults as defined in Humbatova et al.~\cite{Humbatova2020taxonomyofrealfault}. Instead, they aim to improve system performance by introducing new components to be fixed. For example, rather than fixing data or the training algorithm, CARE~\cite{Sun2022Care} and NNRepair~\cite{usman2021nnrepair} attempt to fix model weights. Conventionally, model weights are not considered part of the original machine learning system~\cite{Panichella2021}, as these weights are automatically determined by the training algorithm. Nevertheless, fixing model weights can result in a more secure and performant model~\cite{usman2021nnrepair}. We therefore expand the fault taxonomy~\cite{Humbatova2020taxonomyofrealfault} to accommodate for these cases.

\begin{figure*}
    \centering
    \includegraphics[width=125mm,scale=0.5]{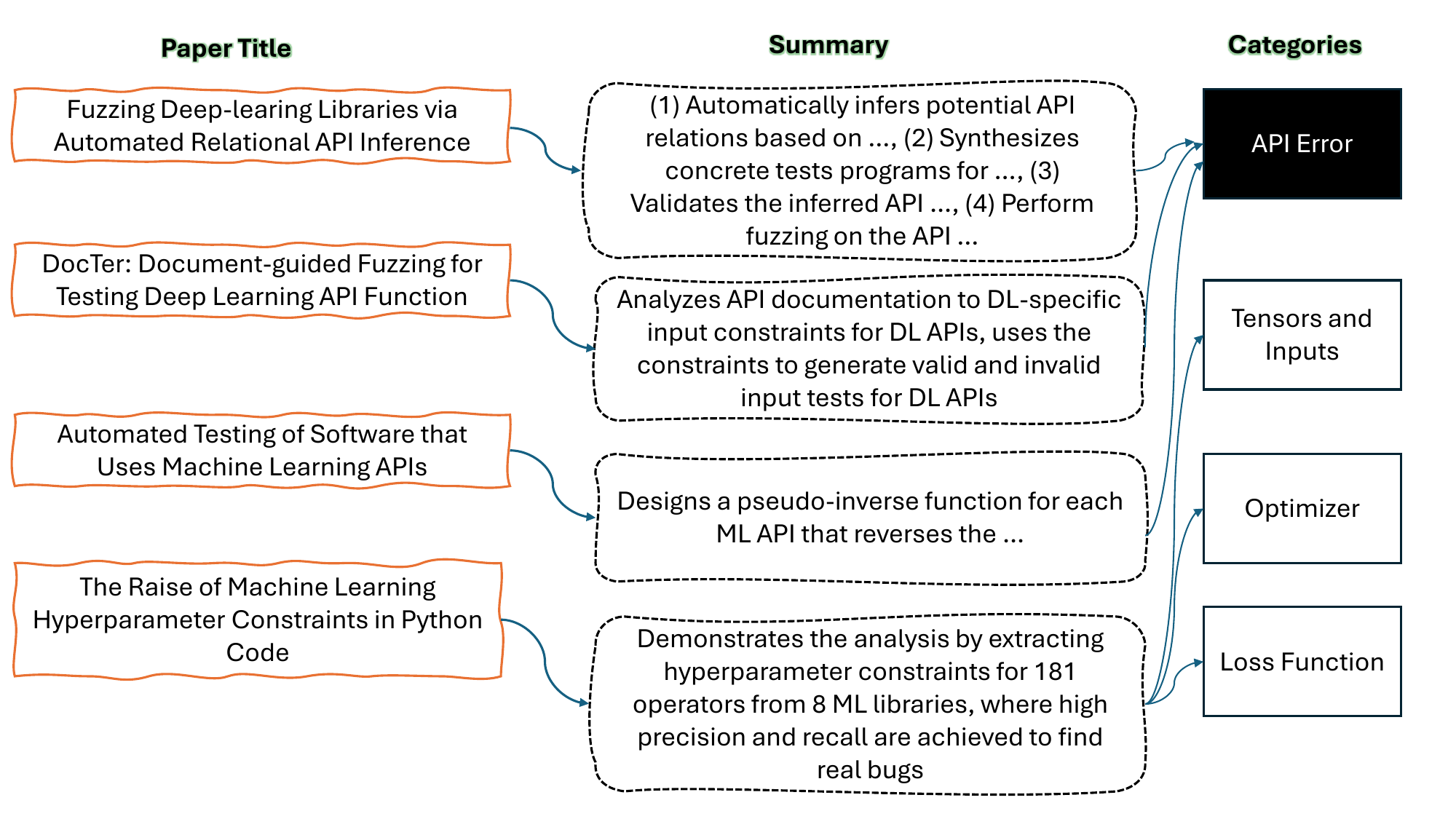}
    \caption{Examples of alignments between debugging techniques and corresponding faults. For each paper, we extract key points where there are mentions of the faults which the paper aims to address and align these faults with the corresponding codes (categories) proposed by Humbatova et al.~\cite{Humbatova2020taxonomyofrealfault}. We expand ~\cite{Humbatova2020taxonomyofrealfault} to accommodate newly emerging categories, where necessary, and denote them with black boxes.} 
    \label{fig:fault_status}
\end{figure*}

\begin{sidewaysfigure*}
    
    \centering    \includegraphics[width=1.05\textheight]{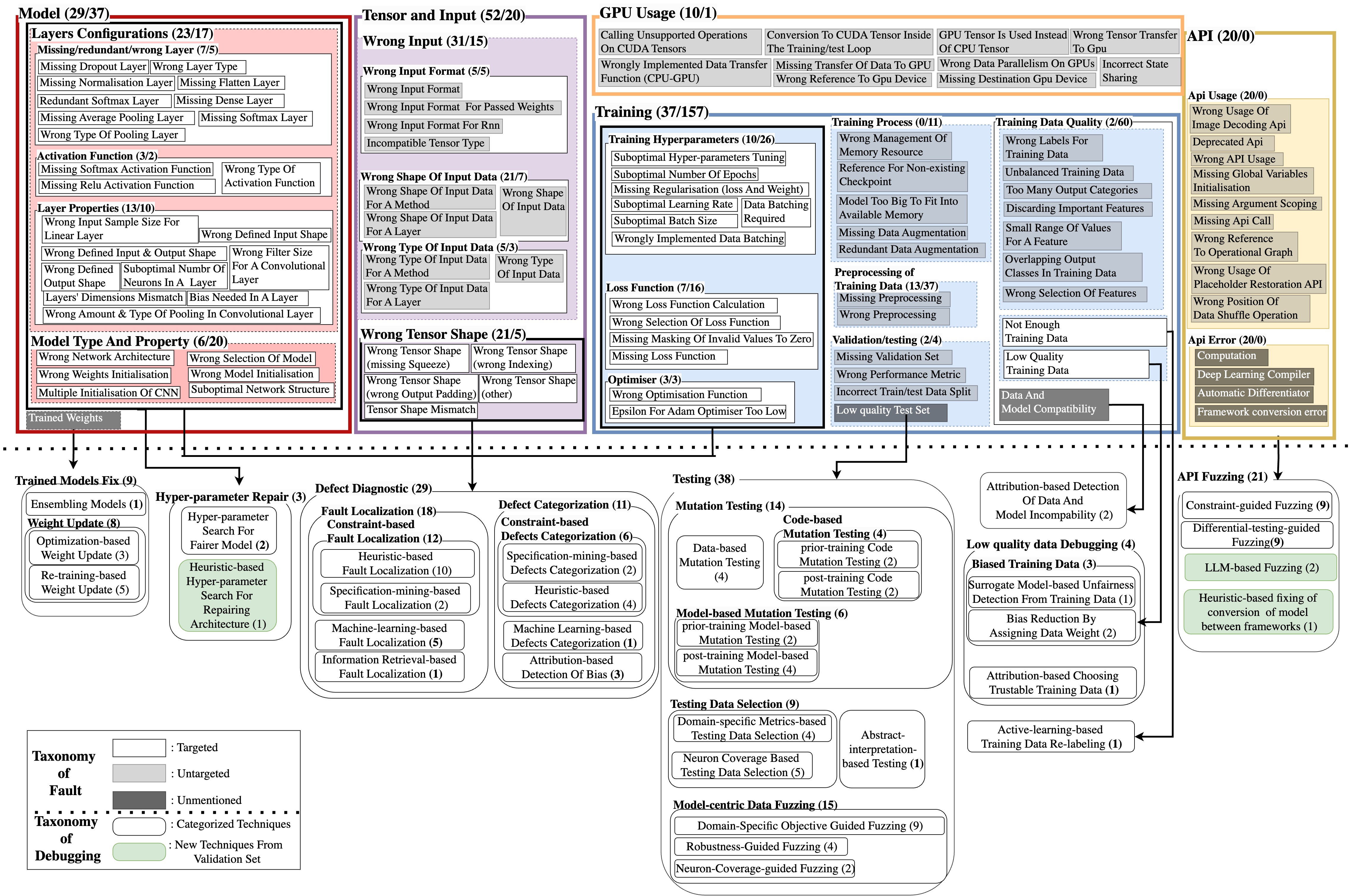}
    \caption{Taxonomy of alignment: The upper part shows the taxonomy of faults with status, and the lower part describes the taxonomy of debugging techniques. The arrows align the faults with the corresponding debugging techniques.}
    \label{fig:taxonomy_of_alignment}
\end{sidewaysfigure*}

We illustrate examples of our mapping process, which links ML debugging techniques to the types of faults they target, in Figure~\ref{fig:fault_status}. Based on the mapping, we categorize the faults in the existing taxonomy into three distinct categories, depicted in Figure~\ref{fig:taxonomy_of_alignment}:

\begin{enumerate}
    \item \textbf{Targeted Faults:} These are faults that directly correspond to those defined in the original taxonomy by Humbatova et al.~\cite{Humbatova2020taxonomyofrealfault}. Papers addressing these faults align closely with the established fault categories.
    
    \item \textbf{Untargeted Faults:} These faults are present in the original taxonomy but are not specifically targeted by any of the collected papers. This indicates areas within the taxonomy that may require further attention or research.
    
    \item \textbf{New Targeted Faults:} These are faults addressed by the papers that were not included in Humbatova et al.'s original taxonomy. These faults represent novel areas of investigation that have emerged in recent research.
\end{enumerate}

Figure~\ref{fig:taxonomy_of_alignment} visualizes the distribution of these categories. The figure plots the ratio of targeted faults versus untargeted faults based on the number of papers that address each category. This visualization highlights the alignment between existing research and the established taxonomy, as well as the emergence of new fault categories that extend beyond the original framework.

\vspace{2mm}\noindent\textbf{Aligning Debugging Techniques and Faults.}
Finally, to align the taxonomy of faults, which we extended from~\cite{Humbatova2020taxonomyofrealfault}, with the taxonomy of ML debugging techniques that we created, we link each debugging technique $d$ to the fault $f$ it addresses. Specifically, for each debugging technique $d$, we identify papers that categorize under $d$ and target fault $f$. For example, the fault \textit{Wrong Tensor Shape}, which describes issues arising from incorrect tensor dimensions in specific layers of a deep learning model or the input data, is addressed by machine learning fault localization methods such as heuristic-based fault localization and specification-mining-based fault localization.

Figure~\ref{fig:taxonomy_of_alignment} serves as a comprehensive visualization of this alignment. The upper part of the figure shows the relationship between faults and the number of papers targeting each fault category, while the lower part details the various debugging techniques and their connections to these faults. This dual representation helps easily understand how current research aligns with existing fault taxonomies and where new areas of debugging techniques are emerging.


\section{Gaps between observed faults and their resolutions}\label{sec5:gaps}
In this section, we explore the existing gaps between the faults identified in machine learning (ML) systems and the corresponding debugging solutions proposed in the literature. This analysis is structured around four primary research questions (RQs) to systematically assess the current ML debugging landscape and identify research areas that require further effort. Our research questions are:
\begin{itemize}
\item \textbf{RQ1.} \textit{Is our construction of the taxonomy of ML debugging comprehensive and reliable?}
\item \textbf{RQ2.} \textit{Which faults in the existing taxonomy of faults \textbf{have been targeted} by ML debugging techniques? Are they indeed being resolved by developers in real-world ML applications?}
\item \textbf{RQ3.} \textit{Which faults in the existing taxonomy of faults \textbf{have not been targeted} by ML debugging techniques and what are the challenges in resolving them?}
\item \textbf{RQ4.} \textit{Ultimately, what are the debugging techniques compiled into our taxonomy of ML debugging?}
\end{itemize}
We start with RQ1, which provides evidences for the comprehensiveness and reliability of our methodology to construct the taxonomy of ML debugging. In RQ2 and RQ3, we explore the alignment between the taxonomy of faults proposed by Humbatova et al.~\cite{Humbatova2020taxonomyofrealfault} and the taxonomy of ML debugging constructed by us. This allows us to understand deeper into how real ML faults have been addressed by research and practice in ML debugging and the remaining challenges lying ahead. Finally, in RQ4, we provide our constructed taxonomy of debugging, which comprehensively reviews papers that we collected via our methodology described in Section~\ref{sec:methodology}. We describe each RQ and answer them in detail in the following.

\subsection{RQ1. The comprehensiveness and reliability of the taxonomy of ML debugging that we constructed.}
As described in the methodology to construct of the taxonomy of ML debugging, we split the set of collected papers to open-coding and validation sets to ensure that our constructed taxonomy cover sufficient topics. That is, we perform coding on the validating set, consisting of papers published from 2023 to 2024 to determine whether the new papers can be categorized into the existing categories and what are the newly emerged categories. We seek to answer the following questions to ensure the comprehensiveness and reliability of our construction method.

\vspace{1mm}\noindent\textbf{How well does the taxonomy from open-coding set capture the new papers?} Out of 34 papers in the validation set, 28 papers can be categorized to the existing categories. The 6 remaining papers constitute 5 new categories that still belong to the higher-order theme. The new categories are: (1) Heuristic-based hyper-parameter search for repairing architecture, (2) Code refactoring for performance optimisation, (3) Similarity-based model prediction correction (4) LLM-based fuzzing, and (5) Heuristic-based fixing of model conversion API between frameworks.
Although it is natural for new research directions to emerge, our taxonomy captures the majority of categories in the validation set.

\vspace{1mm}\noindent\textbf{What are the fatest-growing categories?} We summarize number of papers published in the validation set and identify the most-used techniques. This includes: (1) Domain-specific-objective-guided fuzzing of API, having 6 papers falling in this category, (2) Heuristic-based fault localization having 7 papers, (3) Domain-specific metric-based testing data selection, Constraint-guided fuzzing, and machine learning-based fault localization, having 3 papers each. 
We see that the most focused theme is still API fuzzing, followed by that is applying domain-specific understanding to detect and localize faults.

Now that we have shown evidences that our constructed taxonomy of ML debugging sufficiently covers topics, including those that are new and emerging, we proceed to use the taxonomy to answer next research questions. We refer to details of the taxonomy of ML debugging in Section~\ref{subsec:fix}.

\subsection{RQ2: Faults targeted by ML debugging techniques and their occurrences in real-world ML applications.}
We proceed to use the constructed taxonomy of debugging to study which real ML faults, as categorized by Humbatova et al.~\cite{Humbatova2020taxonomyofrealfault}, have been targeted/untargeted in the taxonomy of debugging. Our analysis results are presented in the \textbf{upper part} of the Figure~\ref{fig:taxonomy_of_alignment}.  

In what follows, we provide: (1) high-level overviews of \textbf{targeted faults, targeted but unmentioned faults, and untargeted faults}, and (2) the alignment of these faults from research to practice by investigating how often they occur or have been resolved on GitHub repositories and in interviews with practitioners. We refer to further details on which debugging techniques address the targeted faults to Section~\ref{subsec:fix}, and which are the untargeted faults and their challenges in Section~\ref{subsec:challenge}.

\begin{tcolorbox}[colframe=white, colback=blue!10, coltitle=black, rounded corners]
RQ2.1 Overviews of faults targeted/untargeted by ML debugging 
\end{tcolorbox}

\vspace{1mm}\noindent\textbf{Targeted Faults.} Illustrated as \textit{white} boxes in the upper part of the Figure~\ref{fig:taxonomy_of_alignment}, these are faults for which at least a ML debugging paper addresses. By following the arrows in Figure~\ref{fig:taxonomy_of_alignment}, the correspoding ML debugging techniques (in the bottom part of Figure~\ref{fig:taxonomy_of_alignment}) that address the faults are shown. According to our analysis, the targeted faults encompass all categories under the \textit{Model Architecture} theme, as well as subsets of the \textit{Tensor and Input} and \textit{Training} themes. This indicates that significant research efforts have been directed toward addressing structural and configuration-related issues within machine learning systems. 
In summary, faults targeted by ML debugging techniques include:

\begin{enumerate}
    \item \textbf{Model Architecture}: Faults arising from mismatches in model design and initialization, such as missing layers or incorrect activation functions.
    \item \textbf{Wrong Tensor Shape}: Errors due to incompatible tensor dimensions required by specific operations or model layers.
    \item \textbf{Training Hyperparameters}: Issues related to external model configurations like learning rate, batch size, number of epochs, and regularization parameters.
    \item \textbf{Loss Functions and Optimizers}: Faults stemming from the incorrect choice or configuration of loss functions and optimizers during training.
    \item \textbf{Data Quality}: Problems related to insufficient or low-quality training data, including mislabeled data and biased datasets.
\end{enumerate}

\vspace{1mm}\noindent\textbf{Targeted but Unmentioned Faults.} Illustrated as \textit{dark gray} boxes in the upper part of the Figure~\ref{fig:taxonomy_of_alignment}, these are faults that have been addressed by machine learning debugging techniques but were not included in the original taxonomy of faults by Humbatova et al.~\cite{Humbatova2020taxonomyofrealfault}. We thus extend the taxonomy of faults proposed by~\cite{Humbatova2020taxonomyofrealfault} to accommodate these unmentioned faults, including:
\begin{enumerate}
    \item \textbf{Trained Weights} – Categorized under the \textit{Model} theme, these faults involve issues related to the weights of trained models, such as incorrect weight initialization or adjustments.
    \item \textbf{Low-Quality Test Set} – Placed under the \textit{Validating/Testing} theme, this fault pertains to the quality and representativeness of the test datasets used to evaluate model performance, including  data duplications, data leakage, or insufficiently covering possible behaviors of neural networks.
    \item \textbf{Data and Model Compatibility} – Classified under the \textit{Training Data Quality} theme, this fault involves mismatches between the input data characteristics and the model’s induction bias, leading to performance degradation. This type of fault can manifest in several ways, including mismatched data types, incorrect input dimensions, or invalid feature representations, leading to suboptimal or erroneous predictions.
    
    \item \textbf{API Error} – Included in the \textit{API} theme, these faults arise from incorrect usage or configuration of machine learning framework APIs, such as deprecated functions or improper parameter settings.
\end{enumerate}



\vspace{1mm}\noindent\textbf{Untargeted Faults.} Illustrated as \textit{gray} boxes in the upper part of the Figure~\ref{fig:taxonomy_of_alignment}, these faults have not been addressed by any existing papers. The untargeted faults include original faults under \textit{GPU Usage} theme, \textit{API} theme, and subsets of the \textit{Tensor and Input} and \textit{Training} themes. The absence of targeted solutions for these faults highlights areas where further research and development are needed to enhance the robustness and reliability of machine learning systems. We investigate the challenge in addressing these faults that have been raised by the practitioners in Section~\ref{subsec:challenge}. 

\begin{tcolorbox}[colframe=white, colback=blue!10, coltitle=black, rounded corners]
RQ2.2 How often those faults addressed by ML debugging research occur in practice?
\end{tcolorbox}

\begin{figure}
    \centering    
    \vspace{-1cm}
    \includegraphics[width=\linewidth]{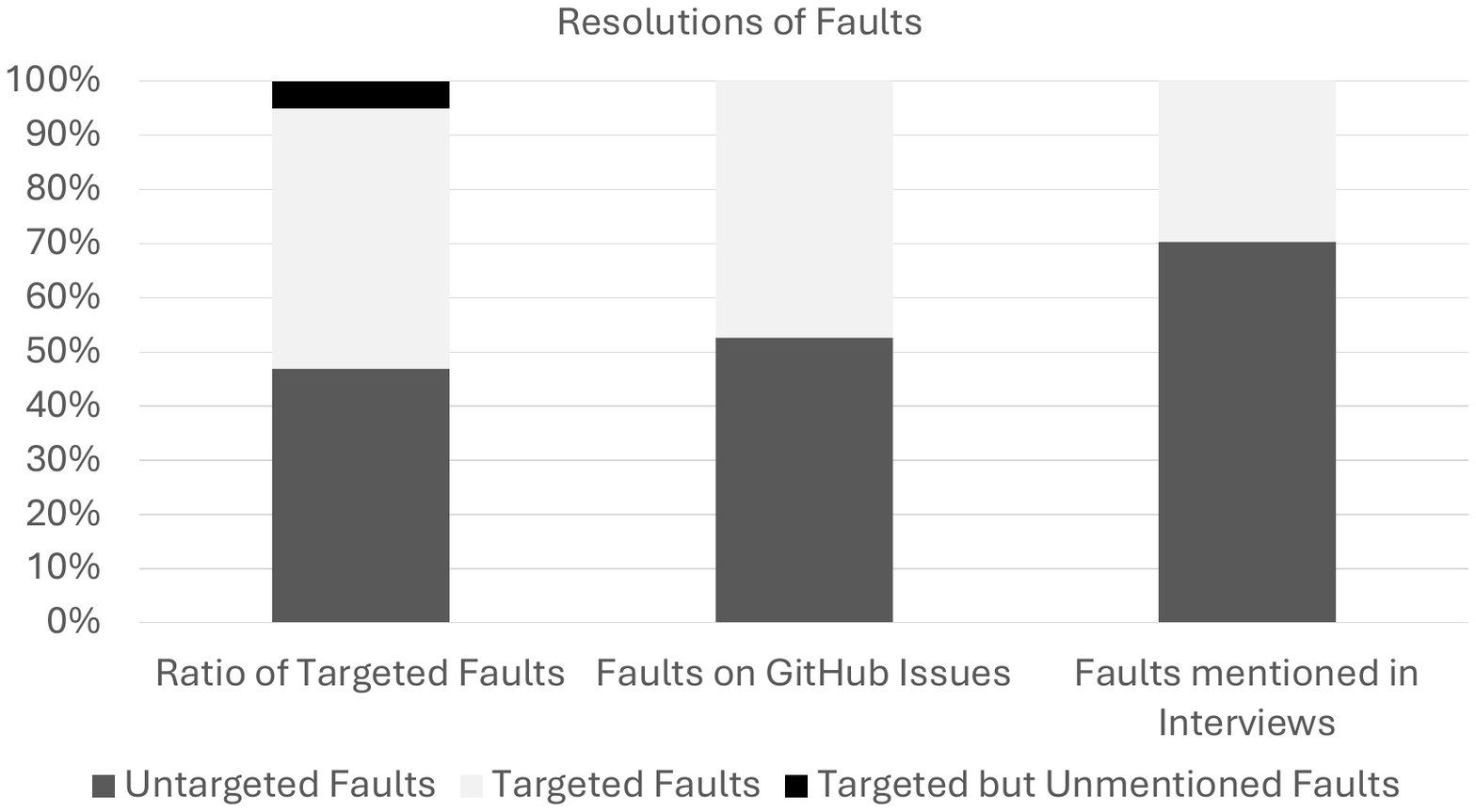}
    \vspace{-1cm}
    \caption{The first column shows the percentage of faults in each category among all faults. The second column represents the ratio in terms of encounters of targeted/untargeted faults on Github issues, and the last column shows the ratio in terms of encounters of targeted/untargeted faults in interviews.}
    \label{fig:ratio}
\end{figure}

We go beyond just simply categorizing the faults by additionally aligning these targeted, untargeted, and unmetioned faults from research to practice to explore how often they occur or have been resolved by developers in real-world ML applications on GitHub and interviews with practitioners. That is,
we count the number times that targeted faults and untargeted faults occur on GitHub issues and interview transcripts provided by~\cite{Humbatova2020taxonomyofrealfault}. The result is shown in Figure~\ref{fig:ratio}. Each bar is divided into three segments (1) Untargeted faults representing faults that have not been targeted by ML debugging techniques. (2) Targeted faults indicating faults that have been targeted by ML debugging techniques. (3) Targeted and unmentioned faults, refering to faults that were targeted by ML debugging techniques but were not mentioned in the original taxonomy of fault by Humbatova et al.~\cite{Humbatova2020taxonomyofrealfault}.

The results in Figure~\ref{fig:ratio} show that there is a high proportion of faults, that are real pressing challenges in practice, are yet untargeted by ML debugging research. Particularly:
\begin{itemize}
    \item \textbf{High proportion of untargeted faults on GitHub and interviews with practitioners.} In both the GitHub and interview cases, the percentage of untargeted faults is relatively high. That is 52.6\% of GitHub issues provided in~\cite{Humbatova2020taxonomyofrealfault} are related to faults that are untargeted by existing ML debugging techniques. Additionaly, 70.3\% of all faults mentioned by practitioners in interview transcripts provided by~\cite{Humbatova2020taxonomyofrealfault} are related to faults that are untargeted by existing ML debugging techniques. This indicates that research efforts in ML debugging has not been sufficiently targeting real pressing challenges and needs in practice. 
    \item \textbf{Nearly half of all faults from the taxonomy of faults have not been targeted.} In the “Ratio of Targeted Faults” (leftmost bar), a substantial portion (46.9\%) of faults from the taxonomy of faults are untargeted by ML debugging techniques, and 48\% have been targeted for resolution. 
\end{itemize}

\begin{figure}
    \centering
    \vspace{-1cm}
    \includegraphics[width=\linewidth]{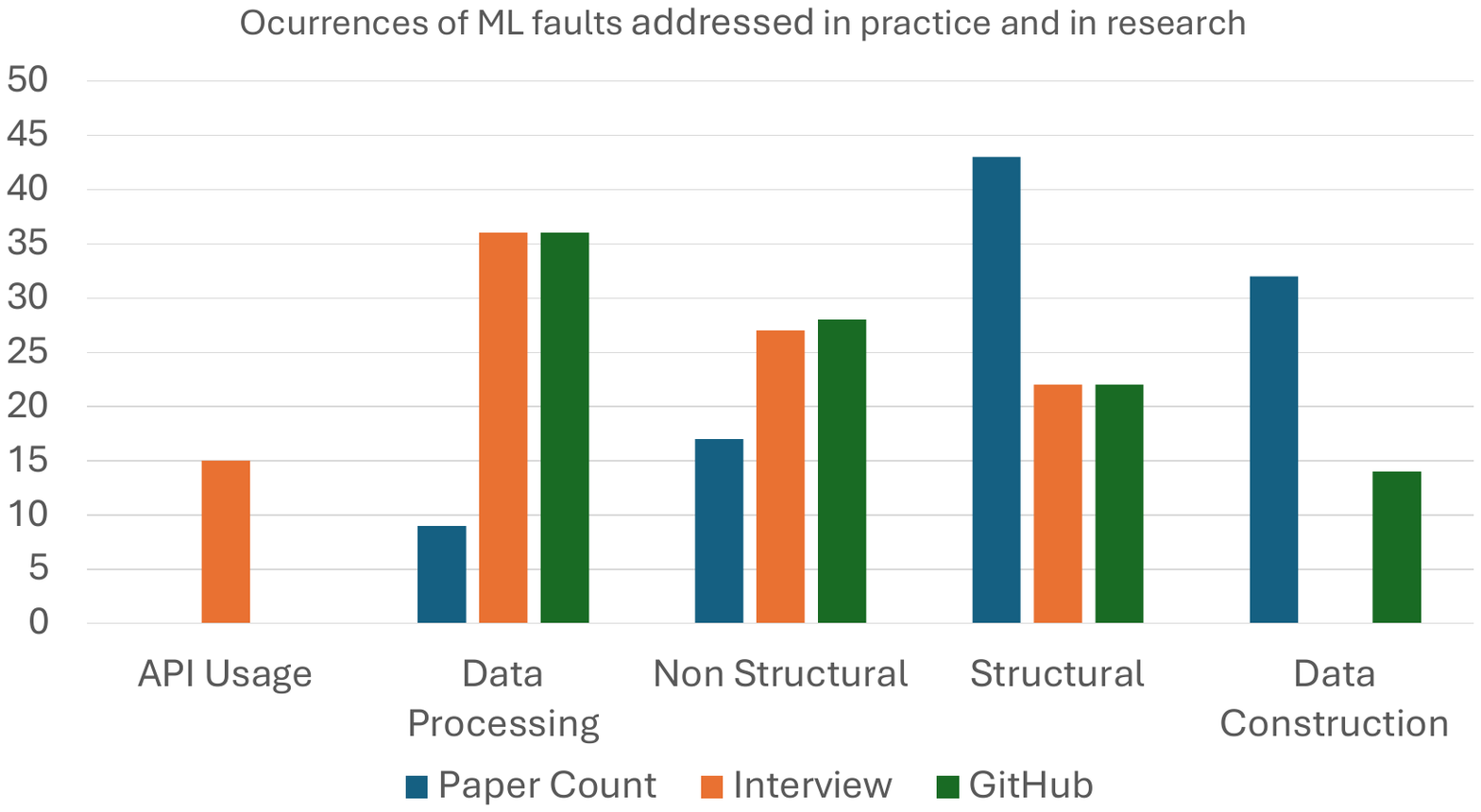}
    \vspace{-1cm}
    \caption{Alignment between ML components addressed/targeted by ML debugging research and the number of observed occurrences those faulty components in Github, and in interviews with practitioners}
    \label{fig:enter-label}
\end{figure}

We show further details on the occurrences of faults targeted by ML debugging research papers and in practice (on GitHub and in interviews with practitioners) in Figure~\ref{fig:enter-label}. To be concise, we merge faults into larger categories such as structural, non-structural, etc, following~\cite{islam2019comprehensivednnbugs}. Notably, faults such as API usage and data processing have been mentioned/encountered quite frequently in practice (in interviews with practitioners and/or in GitHub repositories), but have never been (or very little) addressed by ML debugging research. On the other hand, faults in categories such as structural, non-structural, and data construction have been targeted by ML debugging research, which are in harmony with developers' need in practice. It also shows that more research efforts need to be achieved in the category of faults in data processing to catch up with the high demand of resolving those faults in practice.

\subsection{RQ3: Faults untargeted by ML debugging techniques and the challenges in resolving them.}
\label{subsec:challenge}
In this section, we explain the faults that have not yet been targeted by existing ML debugging techniques (see RQ3.1). More importantly, we identify
the challenges of fixing these faults (see RQ3.2) by studying the interview transcriptions provided by the taxonomy of faults~\cite{Humbatova2020taxonomyofrealfault}, conducting a thematic analysis to discover the common challenges that were mentioned during the interviews.

\begin{tcolorbox}[colframe=white, colback=blue!10, coltitle=black, rounded corners]
RQ3.1. Faults yet to be targeted by ML debugging
\end{tcolorbox}

The untargeted faults are faults that are indicated as gray color in the upper part of the Figure~\ref{fig:taxonomy_of_alignment}. We describe them below.

\vspace{1mm}\noindent\roundbox{\textbf{Wrong Input}} in machine learning systems can manifest in several ways, often leading to catastrophic failures during model training or inference. 

One common issue is \textit{Wrong Input Format}, which occurs when the type of input data (e.g., tensor format) is incorrect or incompatible with the expected format for a specific model architecture, such as Recurrent Neural Networks (RNNs) \cite{islam2019comprehensivednnbugs}. For example, input formats in RNNs need to adhere to specific temporal sequence structures, and any deviation in format can cause the model to fail during training or produce inaccurate predictions.

Another prevalent issue is \textit{Wrong Shape of Input Data}, which happens when the dimensionality of the input does not match the expected shape required by a particular method or layer in the neural network \cite{shapetracer}. For instance, convolutional layers in Convolutional Neural Networks (CNNs) expect 3D input data (height, width, and channels), and providing inputs with the wrong shape will result in immediate runtime errors. While these errors are easily tracked within machine learning systems, the challenge lies in domain-specific data processing, where various tools are used to convert the raw data into tensors.

Additionally, \textit{Wrong Type of Input Data} involves providing data types that are incompatible with the method or layer in use. An example is passing categorical data into a layer expecting numerical inputs, which can disrupt the model’s ability to process data correctly \cite{pham2019cradle}. These input mismatches are particularly critical in deep learning pipelines, where large-scale data preprocessing pipelines are often automated, increasing the likelihood of undetected type mismatches.

\vspace{1mm}\noindent\roundbox{\textbf{Training Process Errors}} are critical to machine learning systems as they directly impact the model’s efficiency and overall performance. Common issues include improper memory resource management, incorrect checkpoints, excessively large models, and data augmentation errors.

One frequent challenge in training machine learning models is the \textit{Wrong Management of Memory Resources}. This occurs when the system does not optimally allocate or release memory during training, which can lead to memory overflow, excessive use of swap space, or reduced training speed. Large models like GPT-3~\cite{brown2020languagemodelsfewshotlearners} or BERT~\cite{devlin-etal-2019-bert}, which require significant memory to store gradients, weights, and intermediate activations, can easily exhaust the available resources on a system with limited memory \cite{patterson2022}. Techniques like gradient checkpointing or using mixed precision can alleviate these issues by optimizing memory usage during the training process \cite{micikevicius2018mixed}.

Another frequent issue is \textit{Reference to Non-Existing Checkpoints}. Checkpoints are crucial for resuming training without starting from scratch. However, errors arise when a training pipeline references a checkpoint file that no longer exists, leading to crashes or faulty results. Ensuring that checkpoint management is robust and that checkpoints are saved at consistent intervals can mitigate this issue \cite{rojas2020checkpointing}.

\textit{Model Too Big to Fit Into Available Memory} refers to the situation when the model architecture is too large for the available GPU or CPU memory. This problem can manifest during training when models attempt to load more data than the system’s memory can handle. Techniques like model parallelism, where parts of a model are distributed across different devices, or reducing batch sizes are common solutions for this issue \cite{you2019largebatch}.

In terms of data handling, \textit{Missing Data Augmentation} and \textit{Redundant Data Augmentation} can significantly affect model performance. Data augmentation is vital for improving a model’s generalization by artificially increasing the diversity of the training data. Missing augmentation can lead to overfitting, as the model might become too specialized for the training data, whereas redundant augmentation can waste computational resources by performing unnecessary transformations repeatedly on the same data samples \cite{yang2023imagedataaugmentationdeep}. Ensuring an efficient and well-balanced data augmentation pipeline helps maintain model robustness without compromising training speed.

\vspace{1mm}\noindent\roundbox{\textbf{Preprocessing of Training Data Errors}} are some of the most common issues in machine learning workflows, as improper handling of data can significantly degrade model performance. Two primary types of errors occur: \textit{Missing Preprocessing} and \textit{Wrong Preprocessing}.

\textit{Missing Preprocessing} happens when critical data preparation steps are skipped, which can result in models receiving raw, noisy, or improperly scaled data. For example, many machine learning models assume that input data is normalized or standardized, and failing to apply these preprocessing steps can lead to poor convergence during training or suboptimal performance \cite{lecun1998efficient}. Another common issue is the absence of handling missing values, which can result in NaNs being fed into models, leading to failed computations or inaccurate predictions~\cite{Emmanuel2021missingdata}. Missing preprocessing is particularly dangerous in real-world data pipelines, where data might be unstructured or incomplete.

\textit{Wrong Preprocessing} refers to applying incorrect transformations or procedures to the training data. For instance, using incompatible data encodings or improperly scaling categorical variables can lead to biases in the model’s training process \cite{kotsiantis2006handling}. Other examples include incorrect data normalization methods, such as standardizing data meant for min-max scaling or vice versa, which can distort model behavior \cite{kuhn2013applied}. The wrong choice of preprocessing can introduce artifacts into the dataset that impair the model’s ability to learn generalizable patterns from the data. Additionally, certain data types, like time-series or text, require specific preprocessing techniques, and applying standard procedures may cause loss of critical information \cite{geron2019hands}.

Addressing these preprocessing errors involves carefully inspecting data pipelines and ensuring that all transformations align with the model’s requirements. It also involves validating that the data fed into the model is both clean and properly prepared, using established best practices for each specific data type.

\vspace{1mm}\noindent\roundbox{\textbf{Validation/Testing Errors}} refer to a series of mistakes that occur during the evaluation stage of machine learning model development, which can significantly impact the reliability and performance of the final model. Common validation/testing errors include \textit{Missing Validation Set}, \textit{Wrong Performance Metric}, and \textit{Incorrect Train/Test Data Split}.

\textit{Missing Validation Set} occurs when no separate dataset is used to validate the model during the training process. This is critical because, without a validation set, there is no intermediate check to ensure the model is generalizing well to unseen data. This omission often leads to overfitting, where the model performs well on the training set but poorly on new, unseen data \cite{hastie2009elements}. Validation sets serve as a critical checkpoint to assess the model’s performance during training, allowing for hyperparameter tuning and model selection.

\textit{Wrong Performance Metric} involves using an inappropriate evaluation metric for the problem at hand. For instance, accuracy might be used as a performance metric in an imbalanced classification problem, where precision, recall, or F1-score would be more appropriate \cite{japkowicz2011evaluating}. This can result in misleading performance evaluations, as certain metrics may mask deficiencies in the model’s ability to handle specific classes or tasks. For example, in a binary classification problem with highly imbalanced classes, a model predicting the majority class could achieve high accuracy while ignoring the minority class entirely.

\textit{Incorrect Train/Test Data Split} is another common issue, where the data is improperly partitioned for training and testing. This includes scenarios where the test set contains data that has already been seen by the model during training, leading to inflated performance metrics. Ideally, the training and testing sets should be mutually exclusive to accurately measure how well the model generalizes to new, unseen data \cite{raschka2018model}. Furthermore, failing to maintain a representative test set that reflects the distribution of real-world data can result in misleading performance results when the model is deployed.

These validation and testing errors are crucial to address for building robust machine learning models, as they directly impact the model’s performance in real-world applications.

\vspace{1mm}\noindent\roundbox{\textbf{Training Data Quality Issues}} involve various faults in the composition and preprocessing of the data used to train machine learning models. These errors can substantially affect the model’s ability to learn patterns and generalize well to unseen data. Some common issues include \textit{Wrong Labels for Training Data}, \textit{Unbalanced Training Data}, \textit{Too Many Output Categories}, \textit{Discarding Important Features}, \textit{Small Ranges of Values for a Feature}, \textit{Overlapping Output Classes in Training Data}, and \textit{Wrong Selection of Features}.

\textit{Wrong Labels for Training Data} refers to situations where the labels (i.e., target variables) are incorrect, leading to misguidance during model training. Models trained on mislabeled data may learn incorrect patterns and produce erroneous predictions. This issue is particularly detrimental in supervised learning, where the model relies heavily on the correctness of the labels to learn \cite{northcutt2011confidentlearning}.

\textit{Unbalanced Training Data} occurs when one or more classes in the dataset are underrepresented compared to others. For instance, in a binary classification task, a model might perform well on the majority class but struggle with the minority class. Techniques such as resampling or cost-sensitive learning are often employed to mitigate this issue \cite{he2009learning}.

\textit{Too Many Output Categories} can lead to performance degradation, particularly in classification problems. If the number of output classes is too large, the model might struggle to distinguish between classes, especially when the classes are similar or there is insufficient data for each category. This can result in higher complexity and lower accuracy \cite{geron2019hands}.

\textit{Discarding Important Features} refers to the omission of features that carry significant predictive power. This can occur during the feature selection process, where key attributes are inadvertently left out, reducing the model’s ability to accurately represent the underlying patterns in the data. Feature importance techniques like SHAP or feature importance scoring methods can help identify and preserve valuable features.

\textit{Small Ranges of Values for a Feature} involves a situation where a feature has a narrow range of values, making it difficult for the model to capture meaningful variations. This is particularly problematic when certain features require normalization or scaling to align with the range of other features in the dataset \cite{deepdiagnosis}.

\textit{Overlapping Output Classes in Training Data} occurs when the classes are not well-separated, leading to confusion for the model. In these cases, the decision boundaries between classes become blurred, which can severely impact the classification performance. This is often addressed through techniques such as dimensionality reduction or cluster-based methods \cite{hastie2009elements}.

\textit{Wrong Selection of Features} can happen when irrelevant or redundant features are selected during the preprocessing phase. This can introduce noise into the model, resulting in overfitting or reduced generalization performance. Feature selection techniques like Lasso regression or principal component analysis (PCA) are commonly used to address this problem \cite{guyon2003introduction}.

\vspace{1mm}\noindent\roundbox{\textbf{API Misuse}} Discrepancies in software systems occur when an API that the software depends on is updated, deprecated, or behaves unexpectedly. These faults are especially prevalent in machine learning (ML) systems due to their rapid development cycles and the frequent breaking changes introduced in libraries and frameworks \cite{zhang2021evolution, Humbatova2020taxonomyofrealfault}. Since machine learning models are highly reliant on APIs for data preprocessing, model training, and deployment, API misuse can result in severe system failures and degraded model performance \cite{huang2023dependency}. Additionally, the semi-automated nature of ML development often involves intricate interactions between APIs, increasing the potential for faults \cite{zhang2021evolution}.

According to Humbatova et al. \cite{Humbatova2020taxonomyofrealfault}, common API usage errors in machine learning systems include: \textit{Wrong Usage of Image Decoding API}, \textit{Wrong API Usage}, \textit{Deprecated API}, \textit{Missing Global Variables Initialization}, \textit{Missing Argument Scoping}, \textit{Wrong Reference to Operational Graph}, \textit{Wrong Position of Data Shuffle Operation}, \textit{Missing API Call}, and \textit{Wrong Usage of Placeholder Restoration API}.

\begin{tcolorbox}[colframe=white, colback=blue!10, coltitle=black, rounded corners]
RQ3.2 Challenges in resolving the untargeted faults\end{tcolorbox}

To better understand the challenges in addressing untargeted issues, we conducted a thematic analysis based on interviews released by Humbatova et al.~\cite{Humbatova2020taxonomyofrealfault}. In this study, we analyzed the key challenges in identifying and resolving problems during the debugging of machine learning programs. From the interview transcripts, we extracted key points related to debugging difficulties. We systematically coded these key points, refining the coding framework after every 30\% of the key points were processed. Ultimately, we grouped these codes into corresponding themes. 

\begin{figure}
    \centering
    \vspace{-1cm}
    \includegraphics[width=\linewidth]{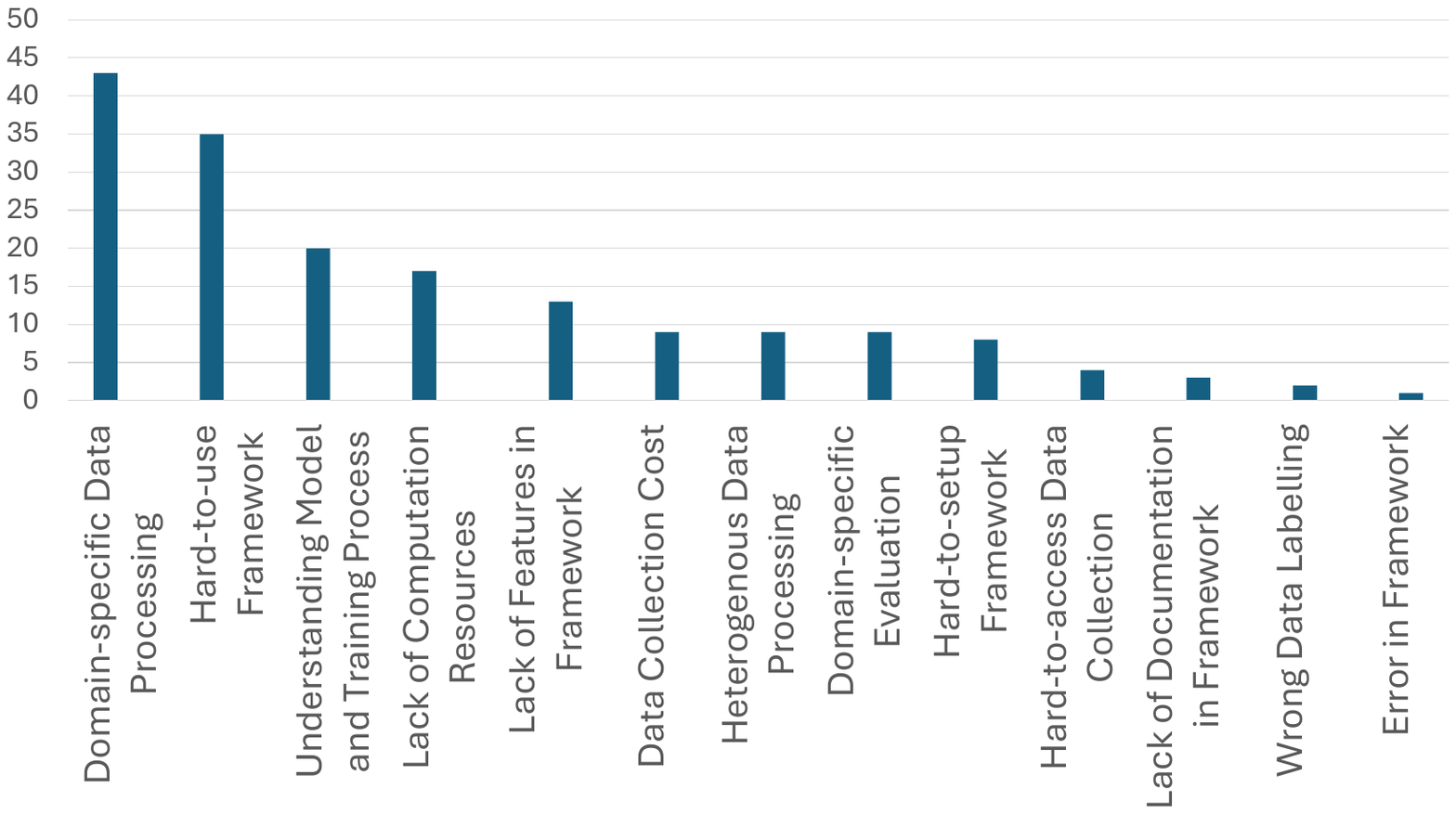}
    \vspace{-1cm}
    \caption{Encounter Frequency for each challenge}
    \label{fig:challenge_freq}
\end{figure}

The most frequent categories are listed in Figure~\ref{fig:challenge_freq}. We identified 13 primary challenges, which we have grouped into the following categories:
\subsubsection{Data-related Challenges}
\begin{enumerate}[label=(\alph*)]
\item \textbf{Domain-specific Data Processing}: ML systems often require specialized tools and techniques for handling domain-specific datasets. This complexity can introduce errors and inconsistencies across different data manipulation tools.
\item \textbf{Data Collection Cost}: Acquiring high-quality, labeled data is often expensive and time-consuming, particularly in domains where data is scarce or difficult to obtain.

\item \textbf{Heterogeneous Data Processing}: Integrating and processing data from multiple sources with varying formats and quality presents significant challenges in ML pipelines.

\item \textbf{Hard-to-access Data Collection}: Legal and regulatory constraints can make it difficult to access necessary data, especially in sensitive domains such as healthcare or finance.

\item \textbf{Wrong Labeling in Data}: Incorrectly labeled training data can lead to suboptimal or faulty models, requiring time-consuming identification and correction processes.
\end{enumerate}

\subsubsection{Framework-related Challenges}
\begin{enumerate}[label=(\alph*)]
\item \textbf{Hard-to-use Framework}: Many ML frameworks provide unclear error messages, making it difficult for developers to diagnose and resolve issues within complex ML pipelines.
\item \textbf{Lack of Features in Framework}: Existing frameworks often lack necessary features for specific debugging tasks, limiting developers' ability to implement specialized fixes or optimizations.

\item \textbf{Hard-to-setup Framework}: The installation and configuration of ML frameworks can be cumbersome due to incomplete installations or missing dependencies.

\item \textbf{Lack of Documentation in Framework}: Insufficient or outdated documentation hinders developers' understanding of framework architecture, API usage, and potential errors.

\item \textbf{Error in Framework}: Bugs within the ML frameworks themselves can impact model training or inference, often requiring developers to wait for updates or patches from framework maintainers.
\end{enumerate}

\subsubsection{Conceptual and Resource-related Challenges}
\begin{enumerate}[label=(\alph*)]
\item \textbf{Understanding Model and Training Process}: Developers, especially those new to ML, often struggle to comprehend model functionality and training processes, leading to poor performance or unrecognized errors.
\item \textbf{Lack of Computational Resources}: Insufficient computational power can cause models to crash, fail to converge, or take excessive time to complete, particularly when working with large datasets or complex models.

\item \textbf{Domain-specific Evaluation}: Standard evaluation methods may be inadequate for assessing model performance in specialized domains, potentially leading to missed critical issues.
\end{enumerate}

\subsection{Taxonomy of ML Debugging Techniques}\label{subsec:fix}
In this section, we provide details of our taxonomy of ML debugging techniques, comprehensively reviewing all papers that we collected following our methodology described in Section~\ref{sec:methodology}. Figure~\ref{fig:taxonomy_of_alignment} visualizes the categories that we constructed for the taxonomy. We describe each category in detail below.

\vspace{1mm}\noindent\roundbox{\textbf{Trained Model Fix}}. In contrast to traditional software, where human developers must modify the source code manually to pass a failing test case, machine learning (ML) models can be retrained on the failing test cases to improve performance without human intervention. This has led to a new line of research focused on leveraging the semi-automated, data-driven nature of ML models to repair them.

\textbf{Re-training-based Weight Update} Re-training-based weight updates focus on identifying samples that can be used to effectively re-training models for better performance. For instance, DeepGini \cite{feng2020deepgini} selects test inputs with low-prediction confidences to fine-tune the model. Similarly, MCP \cite{shen2020mcp} proposes selecting test samples that lie close to the model’s decision boundary to retrain the model.

One of the main challenges in this line of research is minimizing the labeling cost. This challenge aligns well with the active learning methods from the deep learning community. CoreSet \cite{sener2018active} considers the diversity of selected samples and suggests selecting $k$-Centers, where $k$ is based on the labeling budget. Active learning is suitable in these settings because it actively proposes which data to label, although the model itself remains uncontrollable in repair settings. Other approaches include Badge \cite{ash2019deep} and SSL \cite{wang2020consistency}, which address labeling costs by selecting samples with high inconsistencies and using semi-supervised methods to train on both labeled and unlabeled data.

HybridRepair \cite{li2022hybridrepair} selects unlabeled samples that show high similarity to annotated samples, creating a consistency regularization loss to ensure that data with similar predictions have similar representations. They experiment on datasets like CIFAR-10, SVHN, and GTSRB using CNN architectures such as MobileNet, ResNet, and ShuffleNet. They compare their methods with DeepGini \cite{feng2020deepgini}, MCP \cite{shen2020mcp}, CoreSet \cite{sener2018active}, Badge \cite{ash2019deep}, and SSL \cite{wang2020consistency}.

In another approach, DeepState \cite{Liu2022DeepState} focuses on identifying critical test samples that reveal low accuracy in recurrent neural networks (RNNs). They achieve this by creating a stateful metric based on neuron coverage and test their method on RNN architectures such as LSTM, GRU, and BiLSTM.

\textbf{Optimization-based Weight Update} Another direction is directly update the model weight according to specific constraints. For example, causal-based testing and repair techniques leverage structural-causal models (SCM) \cite{Spirtes2001scm} to localize failure weights and search for new weights. The pioneer work CARE \cite{Sun2022Care} has demonstrated that feed-forward neural networks (FFNN) and convolutional neural networks (CNN) can be equivalently represented using SCMs. This representation allowed them to employ Average Causal Effect (ACE) and Conditional Average Treatment Effect (CATE) to perform causal attribution towards hidden neurons. By identifying faulty neurons, they could search for new weights using Particle Swarm Optimization (PSO) to repair the networks. This method was shown to lower both the success rate of backdoor attacks and the violation rate of safety properties in neural networks. CC~\cite{ji2023cc} extends upon CARE\cite{Sun2022Care}, introducing the causality coverage criterion. To facilitate this criterion, they first build a causal graph of the model concerning different datasets. The CC is thus defined as the overlapping between these different causal graphs. The researchers further show that CC can be used to evaluate test data quality, potentially detecting backdoor and adversarial inputs since these input ``produces'' new causal edges.

\textbf{Ensembling Models} Finally, trained models can be combined together to alleviate each individual model's fault: Chen et al.~\cite{Chen2022MAAT} proposes combines multiple trained models that were trained for different purposes (e.g., performance and fairness) by weighting their confidences for best prediction results.



\vspace{1mm}\noindent\roundbox{\textbf{Defect Diagnostic}}. Defect diagnostic consists of two subcategories: defect categorization and fault localization. 

\textbf{Defect Categorization} aims to classify faults into predefined categories based on failed test cases or performance indicators, as outlined by Thung et al. \cite{thung2013}. This process helps identify patterns in defect occurrence, allowing for the development of targeted solutions. Machine learning faults, however, differ from conventional software engineering faults due to the distinct nature of their construction. As a result, defect categories in machine learning differ, with sub-categorization being achieved either through constraint-based approaches or using machine learning models \cite{deepfd}. Most state-of-the-art debugging techniques rely heavily on heuristics \cite{ben2023thedeepchecker, deepdiagnosis, DBLP:conf/icse/ZhangZMS21}.

TheDeepChecker \cite{ben2023thedeepchecker} categorizes deep learning errors into three phases: pre-training, on-training, and post-training defects. Pre-training defects occur before the training process starts and include issues such as imbalanced labels, problems with data distribution, or improper weight initialization. On-training defects arise during the training phase itself, manifesting through problems like non-decreasing loss, gradient instability, performance misalignment, or incorrect parameter updates. Post-training defects occur after training, with issues like distribution shifts from data augmentation or incorrect model behavior during the transition from training to inference. TheDeepChecker detects these conditions by performing distribution measurements on data (input and labels) and model weights for pre-training defects. During training, it monitors loss, gradients, and parameter updates to identify issues such as overfitting or gradient explosion.

DeepDiagnosis \cite{deepdiagnosis} addresses faults through eight key symptoms: dead nodes, saturated activation, exploding tensors, stagnating accuracy and loss, unchanged weights, and exploding or vanishing gradients. Based on these symptoms, DeepDiagnosis categorizes faults into one of several categories: weight initialization, learning rate, improper data, incorrect activation function, and incorrect loss function.

DeepFD \cite{deepfd} monitors a similar set of components but takes a different approach by mutating deep neural network code to introduce common faults. If the performance of the neural network degrades after a mutation, it indicates that the observed fault is likely the cause. These faults are then used to train a fault diagnoser based on machine learning techniques such as KMeans, Decision Tree, or Random Forest.

\vspace{1mm}\textbf{Fault localization} is the process of identifying the specific components in a system that cause performance failures, such as low accuracy or poor convergence during training. Once a fault type has been identified through defect categorization, fault localization aims to pinpoint the exact failing elements within the machine learning system. Existing research focuses on identifying faults in various areas, including model architecture, tensor shape mismatches, improper training hyperparameters, loss functions, optimizers, and deep learning libraries and frameworks \cite{rak2022, deepfd, Kim2022, deepdiagnosis, DBLP:conf/icse/ZhangZMS21, wardat2021deeplocalize, shapetracer, pham2019cradle}. Fault localization in machine learning can be broadly divided into three subcategories: constraint-based, machine-learning-based, and information-retrieval-based fault localization.

Constraint-based fault localization involves identifying failing statements or components by using predefined constraints \cite{shapetracer, rak2022}. For example, Liu et al. \cite{shapetracer} found that shape-related errors account for 63.69\% of bugs collected from Alibaba's PLATFORM-X. To address this, ShapeTracer extends static analysis techniques by capturing shape constraints and resolving them using an SMT solver. This helps in detecting unresolvable constraints that cause errors related to tensor shapes.

Rak et al. \cite{rak2022} take a different approach, focusing on defects in hyperparameters across various machine learning operators. They mine constraints from deep learning library code through a four-step process: (1) identifying exception-raising statements, (2) constructing call graphs, (3) propagating constraints across different methods using these call graphs, and (4) detecting potential API misuses based on the mined constraints.

In DeepDiagnosis \cite{deepdiagnosis}, fault localization relies on categorized defects to guide the identification of faulty components. The system pinpoints issues related to weight initialization, learning rate, activation function, and loss function, enabling the repair of corresponding parts. DeepFD, on the other hand, applies specific templates for each defect category to localize faulty components in the system, making the process more targeted and efficient.

\vspace{1mm}\noindent\roundbox{\textbf{Testing}}
consists of mutation testing, testing data selection, model-centric data fuzzing, and abstract interpretation-based testing

\textit{Test suite quality measurement} There has been work using neuron coverage for testing~\cite{kim2019surprise, ma2018deepgauge, Gao2022adaptivetest}, however, this metric is ineffective: in the work by Chen et al.~\cite{Chen2020DeepNN}, it was found that neuron coverage can reach its maximum in just a few steps. This implies that a single activation can still exhibit different outputs, as demonstrated by DeepGini\cite{feng2020deepgini} and Structural Coverage Criteria for Neural Networks Could Be Misleading \cite{li2020structuralcoverage}. DeepXplore~\cite{pei2017deepxplore}, leverages neuron coverage and differential testing to identify failing behavior of neural networks. It formulates the problem of finding an input that satisfies domain constraints while maximizing both neuron coverage and differences between a set of neural networks. Using a meta-gradient optimization, DeepXplore found inputs that improved performance in autonomous driving tasks by up to 3\%. Similarly, DeepTest~\cite{Tian2018} explored inputs that maximized neuron coverage and identified erroneous behavior in models for autonomous driving, improving DNN performance in the Udacity self-driving challenge by 46\%. TensorFuzz~\cite{odena2019tensorfuzz} (PRML’19) combined neuron coverage with mutation testing in the domain of image processing. It generated inputs by mutating data (e.g., adding white noise) and identifying numerical bugs like NaN in quantized neural networks, outperforming random search methods in identifying issues in models like those trained on MNIST.
Recent work by Yuan et al. \cite{yuan2023nlc} revised neuron coverage criteria and proposed a layer-wise, distribution-aware criterion called NeuraL Coverage (NLC). They argue that DNNs approximate distributions using hierarchical layers, and test sets should diversify these approximations to assess their quality. NLC captures key properties of neuron output distributions, providing a more robust estimate of test suite diversity. Their results show that NLC correlates strongly with test suite diversity and error discovery, outperforming traditional neuron coverage criteria in multiple tasks.

\textit{Mutation Testing to Measure Test Data Quality} Another direction in neural network testing is mutation testing. Classical mutation testing evaluates test set quality by generating program variants (mutants) and checking if the test cases can detect differences between the original and mutated programs. 
Mutation testing on machine learning also aims to evaluate the test set quality. They do it by mutating either: (1) the networks~\cite{Shen2018Munn, ma2018mutation, hu2019deepmutationplus},  (2) the code implementing the network, the data, and the training procedure \cite{ma2018mutation}, \cite{hu2019deepmutationplus}, \cite{humbatova21deepcrime}.
Due to the aforementioned differences between producing a machine learning model and producing human codes, different settings lead to different approaches.
On this end, Shen et al.\cite{Shen2018Munn} and Ma et al.\cite{ma2018mutation} are the first to propose mutation testing on neural networks. As one of the first studies, MuNN~\cite{Shen2018Munn}'s authors consider a trained CNN model equivalent to the production code in classic mutation, and analyze the following question: \textit{given a popular well-known test set, how do model-centric mutations affect the model's performance on this test set?}
In another sense, this question can be paraphrased as \textit{if we consider the test dataset equivalent to the test suite in classic mutation settings, and whether the test dataset \emph{kills} the mutant depends on whether the test results deviate significantly given the mutated models, will mutation testing works?} 
Given this formulation, they apply five mutation operators to modify neurons, activation functions, bias, and weight values on the trained model to evaluate the deviation in terms of test set accuracy. Their results suggest that mutations can be killed by the test suite, and suggest that domain-specific mutations can be developed for more effective test adequacy measurement.
Ma et al.~\cite{ma2018mutation}, on the other hand, proposes that we regard the whole model production process (i.e., both the data, pre-training, training, and trained model) as production code to be tested. Thus, they propose DeepMutation~\cite{ma2018mutation}, a framework for mutation testing of deep neural networks training process. DeepMutation~\cite{ma2018mutation} and DeepMutation++~\cite{hu2019deepmutationplus} perform mutating both the data by duplicating data, adding labeling errors, deleting, and adding noise to the training data. For models-centric mutations, they perform removing layers, adding layers, and removing activation functions. 
DeepCrime~\cite{humbatova21deepcrime} further improves mutation testing using real-faults by Humbatova et al.~\cite{Humbatova2020taxonomyofrealfault} as well as proposing heuristics to choose suitable mutations based on domain-specific conditions, showing that the chosen mutations have higher sensitivity towards the test set in comparison with DeepMutation++~\cite{hu2019deepmutationplus}.

The challenges of applying mutation testing to machine learning models revolve around issues such as low-quality mutants, the potential use of mutants for retraining, and uncertainties regarding the accuracy of mutation-based metrics. Low-quality mutants, which do not meaningfully impact the model, can distort the results of mutation analysis, making it difficult to determine if the test set or the model is genuinely robust. Additionally, while retraining a model on mutants could be beneficial, it’s often unclear whether such retraining would improve generalization or simply overfit to the mutated examples, reducing the overall utility of mutation testing.

Furthermore, mutation testing in machine learning is closely related to adversarial detection and metamorphic testing. Just as traditional software testing identifies edge cases that reveal vulnerabilities, adversarial attacks on machine learning models exploit subtle manipulations in input data to deceive the model into making incorrect predictions \cite{goodfellow2015explainingharnessingadversarialexamples, carlini2017towards}.

Metamorphic testing, a software testing technique where outputs from modified inputs are compared to detect inconsistencies, can be applied in machine learning to identify such adversarial weaknesses. By modifying input data, for example, through adversarial perturbations, and comparing the resulting outputs, vulnerabilities in the decision-making processes of machine learning models can be detected~\cite{rehman2023metamorphicml, besold2017neural}. This approach helps in identifying models that fail to maintain consistency in their predictions under slight modifications to the input, making it a valuable tool for improving robustness and security in machine learning systems.

\vspace{1mm}\noindent\roundbox{\textbf{Low-Quality Training Data and Insufficient Data}} are critical factors that can severely impact a model's performance and generalization. Low-quality data, such as mislabeled instances, unbalanced class distributions, or irrelevant features, often lead to model biases or reduced accuracy. These issues are difficult to diagnose due to the black-box nature of machine learning models, and can be mistakenly attributed to architecture or hyperparameters.

Unbalanced datasets may cause the model to overfit to dominant classes, neglecting others, while mislabeled data introduces incorrect learning patterns that manifest in poor validation and testing outcomes. Insufficient data, on the other hand, increases the risk of overfitting, where the model performs well on training data but generalizes poorly to new inputs. Inadequate training data often results in high variance in performance, as the model fails to capture meaningful patterns in the data, making it sensitive to outliers and prone to errors during deployment. 

While the problem of low-quality training data is challenging, researchers have started adopting techniques from active learning to reduce the cost in annotating new data~\cite{li2022hybridrepair}, choosing a trusted high-quality set~\cite{zhang2018trainingsetdebugging}, or detecting and mitigate low quality bias training data using surrogate model~\cite{Li2022ltdd}, by assigning data weight. For low quality data, there exist an attribution-based method to choose trustable training data~\cite{zhang2018trainingsetdebugging}.
Zhang et al.~\cite{zhang2018trainingsetdebugging} proposes a method called DUTI (Debugging Using Trusted Items) for identifying bugs in machine learning training sets, particularly mislabeled data. This method aims to fix potential training set issues by leveraging a small number of verified “trusted items” that are known to be correctly labeled by domain experts. The hypothesis is that the training data set would contain mislabeled items or systematic biases that negatively impact the performance of machine learning models, manually verifying and correcting all training data is not feasible for large datasets, but verifying a small number of trusted items is possible. Thus, DUTI seeks to find the minimal set of label changes in the training data so that when a model is retrained on the modified data, it predicts the trusted items correctly. The flagged items are suspected as mislabeled and are given to human experts for further validation. To do this, DUTI transforms the debugging problem into a bilevel optimization problem, where the upper-level optimization seeks to minimize the difference between the original and corrected labels in the training set while ensuring the model’s predictions align with the trusted items, while the lower-level optimization focuses on retraining the model using the corrected labels to fit both the training set and the trusted items. DUTI can be applied to both regression and classification problems. In experiments, the method demonstrated effectiveness in identifying bugs and suggesting label fixes on synthetic and real datasets.

\vspace{1mm}{\noindent\roundbox{\textbf{Hyper-parameter Repair}}. Dutta et al.~\cite{10.1145/3460319.3464844} propose an approach to choose optimized hyper-parameters in such a way that the test execution of ML software gets faster while still preserving desired correctness. To find the optimized version of the
tests, the approach systematically navigates the trade-off space between execution time of each test and its passing probability by tuning the algorithm hyper-parameters. Niari et al.~\cite{10.1145/3510003.3510202} address fairness by the selection of hyperparameters, which provide
finer controls of ML algorithms. They proposed search-based software testing algorithms to uncover the precision-fairness frontier of the hyper-parameter space, successfully identifying hyper-parameters that significantly improve the fairness without sacrificing precision.

\vspace{1mm}\noindent\roundbox{\textbf{Data-and-model incompatibility}}. Yona et al.~\cite{Yona2021algorithmdata} use Shapley value, a concept from cooperative game theories that is used to measure the contribution of different players in cooperative settings, to measure the contribution of model or data to the model performance.

\vspace{1mm}\noindent\roundbox{\textbf{API Fuzzing}}. API fuzzing is a prominent technique in testing deep learning frameworks, relying on approaches like constraint-guided and differential-guided fuzzing. Fuzzing is a software testing method that subjects a program to random, invalid, or unexpected inputs in order to identify vulnerabilities such as program crashes, memory leaks, and unhandled exceptions. In the context of deep learning frameworks, fuzzing has become a key tool for identifying bugs, especially in their APIs.

\textbf{Differential Fuzzing} Differential testing refers to identifying testi input that leads to discrepancies between multiple different versions or implementations. For example, Muffin\cite{Gu2022muffin}, applies differential testing by generating neural models and training them to detect inconsistencies across different versions of the same deep learning framework. Similarly, FreeFuzz\cite{wei2022freefuzz} collects code and model instrumentation and performs mutation-based fuzzing on API inputs. By leveraging oracles based on differential testing and metamorphic testing, FreeFuzz identifies computational errors, crash bugs, and performance issues, comparing different backends such as CPUs and GPUs.

DeepREL~\cite{Deng2022DeepRel}, constructs a large dataset of corresponding API pairs between major deep learning frameworks like PyTorch and TensorFlow. Through fuzzing, it identifies mismatches between API outputs, with these inconsistencies being regarded as bugs. This approach demonstrated significant improvements over FreeFuzz in terms of API output accuracy and bug detection.

In another approach, NablaFuzz~\cite{Yang2023NablaFuzz}, focuses specifically on testing the automatic differentiation components of deep learning libraries. It uses fuzzing and differential testing to examine consistency in gradient computations across frameworks like PyTorch, TensorFlow, JAX, and OneFlow. NablaFuzz generates forward tensors, calculates reverse gradients, and checks for inconsistencies between repeated calls, as well as between calculated and numerically differentiated gradients. This method has been shown to outperform other state-of-the-art fuzzers, including Muffin and FreeFuzz, in terms of both code coverage and bug detection.

\textbf{Constraint-guided Fuzzing}. This category refers to techniques that uses contraints, e.g., input contraints such as typing information, etc, to help guide fuzzing. Xie et al.~\cite{xie2022docter} propose \textit{DocTer}, an approach that leverages documentations, e.g., free-form API documentations, to generate specific input constraints for deep learning libraries. Based on these extracted input constraints such as input types (e.g., list, map, etc) and their properties, \textit{DocTer} generates valid inputs that can test deeper functionalities of deep learning libraries. Wang et al.~\cite{wang2020deep} propose \textit{LEMON}, an approach that tests deep learning models using mutation operators to mutate models. Mutations are designed to explore unused library code and various invoking sequences of library code (APIs). The approach constrains the search space by preferring mutants that have larger inconsistencies compared to the original models. In a similar vein, Guo et al.~\cite{guo2020audee} propose \textit{Audee} to test deep learning frameworks by adopting a search-based approach that implements three different mutation strategies to generate diverse test cases for model structures, parameters, weights and inputs. Wan et al.~\cite{wan2022automated} propose an approach to test software which uses cognitive ML APIs, e.g., image search or classification. It does so by designing a set of pseudo-inverse functions for cognitive ML
APIs, and using symbolic execution to generate tests satisfying the inverse functions. The generated tests can then be used as test inputs for the original APIs.
Yan et al.1\cite{yan2021exposing} propose \textit{GRIST}, an approach to generate failure-inducing inputs for deep learning programs, including training samples and external values (e.g., those produced by random number generators). It uses a static analysis technique to identify potential operations that are likely related to numerical exceptions, and then uses dynamic anlysis to effectively generate tests for those. Shi et al.~\cite{shi2023acetest} propose a technique to automatically extract input validation constraints from the code
and use them to create diverse test cases for core function logic of DL operators. Ma et al.~\cite{ma2023fuzzing} propose an automated testing technique that effectively exposes coding mistakes in deep learning compilers, in the optimization
of high-level intermediate representations (IRs). It does so by employing metamorphic and differential testing while retaining integrity constraints, such as type match and tensor
shape match, that govern high-level IRs to avoid an early crash
before invoking optimization. Wang et al.~\cite{wang2023gencog} propose \textit{GenCoG}
that uses a domain-specific language for specifying type constraints of operators, concolically solves the constraints, and incrementally generates computation graphs of high expressivity to test deep learning compilers. Wei et al.~\cite{wei2022freefuzz} propose \textit{FreeFuzz}, as described in the differential fuzzing category. \textit{FreeFuzz} improves upon its counterpart \textit{LEMON}~\cite{wang2020deep}, performing mutation-based fuzzing on inputs of deep learning APIs. 


\section{Implications for Research and Framework Developers in ML Debugging}\label{sec6:implications}
Based on the comprehensive analysis of challenges in debugging machine learning systems, we provide several key implications emerge for both researchers and framework developers. These implications highlight areas for future research, development, and improvement in the field of ML debugging.
\subsection{Research Implications}
\subsubsection{Advancing Debugging Techniques for Complex Data Processing}
Research should focus on developing more sophisticated debugging tools that can handle domain-specific and heterogeneous data processing. This includes:
\begin{itemize}
\item Creating methods to trace data transformations across different tools and libraries
\item Developing visualization techniques for complex data pipelines
\item Investigating automated error detection in data preprocessing steps
\end{itemize}
\subsubsection{Improving Model Interpretability}
To address the challenge of understanding model and training processes, research should prioritize:
\begin{itemize}
\item Developing more intuitive visualization tools for model architectures and training dynamics
\item Advancing explainable AI techniques to make model decisions more transparent
\item Creating interactive debugging environments that allow step-by-step inspection of model behavior
\end{itemize}
\subsubsection{Enhancing Test Suite Quality Measurement}
Given the limitations of current testing methods, research should focus on:
\begin{itemize}
\item Developing more robust metrics beyond neuron coverage for assessing test suite quality
\item Investigating the relationship between test suite diversity and model robustness
\item Creating tools for automatic generation of high-quality, diverse test sets
\end{itemize}
\subsubsection{Advancing Mutation Testing for ML Models}
To improve the effectiveness of mutation testing in ML, research should explore:
\begin{itemize}
\item Developing domain-specific mutation operators that better reflect real-world ML faults
\item Investigating the relationship between mutants and model generalization
\item Creating metrics to assess the quality and usefulness of generated mutants
\end{itemize}
\subsubsection{Addressing Data Quality and Bias}
Research efforts should be directed towards:
\begin{itemize}
\item Developing automated techniques for detecting and correcting mislabeled data
\item Creating methods to identify and mitigate bias in training datasets
\item Investigating the impact of data quality on model performance and fairness
\end{itemize}
\subsection{Implications for Framework Developers}
\subsubsection{Improving Framework Usability}
To address the challenges related to hard-to-use and hard-to-setup frameworks, developers should focus on:
\begin{itemize}
\item Providing clearer, more informative error messages
\item Developing user-friendly interfaces for framework configuration and usage
\item Creating comprehensive, up-to-date documentation with practical examples
\end{itemize}
\subsubsection{Enhancing Debugging Features}
Framework developers should prioritize the integration of advanced debugging tools, including:
\begin{itemize}
\item Built-in profiling and visualization tools for model training and inference
\item Integrated testing frameworks with support for mutation testing and metamorphic testing
\item Automated suggestions for hyperparameter tuning and model architecture improvements
\end{itemize}
\subsubsection{Improving Framework Robustness}
To reduce errors within the frameworks themselves, developers should:
\begin{itemize}
\item Implement more rigorous testing procedures, including extensive API fuzzing
\item Develop automated compatibility checks for different versions and backends
\item Create tools for easy reporting and tracking of framework-related bugs
\end{itemize}
\subsubsection{Supporting Heterogeneous Computing Environments}
Given the challenges related to computational resources, framework developers should:
\begin{itemize}
\item Improve support for distributed and cloud-based training
\item Develop more efficient memory management techniques
\item Create tools for easy scaling of models across different hardware configurations
\end{itemize}
\subsubsection{Enhancing Support for Domain-Specific Applications}
To address challenges in domain-specific data processing and evaluation, frameworks should:
\begin{itemize}
\item Provide extensible interfaces for integrating domain-specific data processing tools
\item Develop libraries of domain-specific evaluation metrics and testing procedures
\item Create specialized modules for common domain-specific tasks (e.g., healthcare, finance)
\end{itemize}
\subsection{Collaborative Opportunities}
\subsubsection{Standardization Efforts}
Researchers and framework developers should collaborate on:
\begin{itemize}
\item Developing standardized benchmarks for assessing ML debugging tools
\item Creating common interfaces for integrating third-party debugging tools with ML frameworks
\item Establishing best practices for ML system development and debugging
\end{itemize}

\section{Conclusion}\label{sec7:conclusion}
The field of machine learning debugging presents unique challenges that extend beyond traditional software debugging practices. This study has highlighted several critical areas where both researchers and framework developers can contribute to improving the reliability, efficiency, and transparency of ML systems.

Our analysis revealed that: (1) A significant proportion of ML debugging issues remain unresolved, particularly those identified in interview settings, suggesting a gap between theoretical understanding and practical implementation. (2) The challenges in ML debugging span across multiple domains, including data processing, framework usability, model interpretability, and resource management. (3) Current debugging techniques, while promising, are often insufficient to address the complex nature of ML systems, especially in domain-specific applications.

The complexity of ML systems necessitates a paradigm shift in how we approach debugging. By addressing the challenges and implications outlined in this study, the ML community can work towards creating more robust, interpretable, and reliable systems. This not only benefits developers and researchers but also contributes to the broader goal of building trustworthy AI systems that can be confidently deployed in critical applications.

\bibliographystyle{IEEEtran}
\bibliography{sample}

\end{document}